\documentclass[11pt,letterpaper]{article}
\usepackage{soul}
\usepackage{jheppub}
\usepackage{extarrows}
\usepackage[mathletters]{ucs}
\usepackage[utf8]{inputenc}
\usepackage[T1]{fontenc}
\usepackage{amsfonts}
\usepackage{amsmath}
\usepackage{graphicx,subfigure}
\usepackage{tensor} 
\usepackage{mathtools} %
\allowdisplaybreaks[4]
\usepackage{amssymb}
\usepackage{comment}
\usepackage{amsthm,amsmath,amssymb}
\usepackage{mathrsfs}
\usepackage{euscript}
\usepackage{color}
\usepackage{braket}
\usepackage{orcidlink}

\title{Correlators in $T\bar{T}$ and Root-$T\bar{T}$ Deformed CFTs}
\author{Bo-Rui Li$^{a,~b,~}$\footnote{320220902891@lzu.edu.cn},
Song He$^{c,~}$\footnote{hesong@nbu.edu.cn, Corresponding author}, Yu-Xiao Liu$^{a,~b,~}$\footnote{liuyx@lzu.edu.cn, Corresponding author}}

\affiliation{$^{a}$Key Laboratory of Quantum Theory and Applications of MoE, Lanzhou Center for Theoretical Physics,
Key Laboratory of Theoretical Physics of Gansu Province, Gansu Provincial Research Center for Basic Disciplines of Quantum Physics, Lanzhou University, Lanzhou 730000, China\\
$^{b}$Institute of Theoretical Physics $\&$ Research Center of Gravitation, School of Physical Science and Technology, Lanzhou University, Lanzhou 730000, China \\$^c$
Institute of Fundamental Physics and Quantum Technology, \& School of Physical Science and Technology, Ningbo University, Ningbo, Zhejiang 315211, China}

\abstract{
Quasi-primary correlators in two-dimensional conformal field theories deformed simultaneously by $T\bar T$ and root-$T\bar T$ are studied. A path-integral formulation motivated by the geometric realization of the combined deformation is used to develop a geometric framework for evaluating the deformed correlators. Within this framework, the two-point function is obtained to all orders in the $T\bar T$ coupling and to leading order in the root-$T\bar T$ coupling, while the leading correction to the three-point function is computed. It is further shown that the deformed two-point correlator admits a kernel representation as a weighted average of undeformed CFT correlators over conformal dimensions. This representation is derived explicitly for both the pure $T\bar T$ deformation and the combined flow. In this way, the mixed $T\bar T$/root-$T\bar T$ deformation is incorporated into the geometric description of irrelevant deformations, and the structure of local correlators beyond the pure $T\bar T$ case is characterized more explicitly.
}

\begin{document}

\maketitle

\section{Introduction}
An irrelevant composite operator in two-dimensional quantum field theories, denoted by $T\bar T$, was introduced by Zamolodchikov in \cite{Zamolodchikov:2004ce}. In Euclidean signature, it is defined by
\begin{equation}
O_{T\bar T}:= -\det(T^\mu{}_{\nu}) \, .
\end{equation}
Under general assumptions, including translational and rotational invariance, the existence of a conserved stress tensor, factorization at large separation, and a short-distance conformal fixed point, its expectation value satisfies the universal relation
\begin{equation}
\langle O_{T\bar T} \rangle = -\frac{\pi^2}{4}\langle T^\mu{}_{\mu}\rangle^2 \, .
\end{equation}
This factorization property underlies the solvable character of the associated deformation and motivates the study of observables away from the undeformed conformal point. Correlation functions can be used to examine how locality, operator mixing, and short-distance singularities are modified by the deformation. For reviews, see~\cite{Monica,Jiang:2019epa,He:2025ppz}.

At the classical level, starting from an undeformed action $S_0$, the $T\bar T$ flow is defined by
\begin{equation}\label{TTbar Def}
\left\{
\begin{aligned}
S^{(0)} &= S_0,\\
\frac{\partial S^{(\lambda)}}{\partial \lambda}
&= \int_M \mathrm{d}^2x\,\sqrt{|g|}\, O_{T\bar T}^{(\lambda)} \, ,
\end{aligned}
\right.
\end{equation}
where $\lambda$ has dimension $[\mathrm{length}]^2$, $(M,g)$ is a two-dimensional manifold, and $g$ denotes the determinant of the background metric. This deformation preserves integrability in many examples~\cite{Smirnov:2016lqw,Cavaglia:2016oda,Guica:2017lia,Rosenhaus:2019utc,LeFloch:2019wlf,Jorjadze:2020ili}, and it also relates a number of otherwise distinct models~\cite{Cavaglia:2016oda,Bonelli:2018kik,Conti:2018jho}. In addition, the close relation between $T\bar T$ deformations and two-dimensional gravity provides a non-perturbative perspective on the deformation~\cite{Dubovsky:2017cnj, Dubovsky:2018bmo, Tolley:2019nmm}; see also \cite{Nix:2025plr} for applications. More specifically, on curved backgrounds, the classical $T\bar T$ deformation is equivalent to coupling the undeformed theory to a two-dimensional ghost-free massive gravity theory~\cite{JACKIW1985343,deRham:2010kj, Tolley:2019nmm}. In this formulation,
\begin{equation}
S_{T\bar T\text{-deformed QFT}}(f,\phi)
=
S_{\text{QFT}}(e^*(f),\phi)+S_{\text{grav}}(e^*(f),f),
\end{equation}
where $e$ and $f$ are the frame fields of the undeformed and deformed geometries, respectively, $\phi$ collectively denotes the matter fields, and $e^*(f)$ is determined by
\begin{equation}\label{saddle point eq}
\left.
\frac{\delta}{\delta e^a_{\mu}}
\bigl(
S_{\text{grav}}(e,f)+S_{\text{QFT}}(e,\phi)
\bigr)
\right|_{e=e^*(f)}
=0 \, .
\end{equation}

Meanwhile, many global properties of the deformed theory can be treated analytically. At the quantum level, analyzing local correlators involves additional difficulties, since the deformation is irrelevant and the resulting theory is non-local at short distances. A result in this direction was obtained in the large-$c$ analysis of~\cite{Aharony:2018bad}, where stress-tensor correlators were shown to be computable in an appropriate 't~Hooft-like limit, and the deformed three-point function of the stress tensor was derived explicitly. The diffusion equations satisfied by deformed partition functions were realized in a JT-gravity path-integral framework in~\cite{Dubovsky:2018bmo}, and were subsequently studied from the viewpoint of random geometry on different topologies in~\cite{Cardy:2018sdv}. On the torus,~\cite{Datta:2018thy} showed that the partition function of a $T\bar{T}$-deformed CFT on $\mathbb{T}^2$ exhibits $\mathrm{SL}(2,\mathbb{Z})$ modular invariance, while related residual modular properties were further analyzed in~\cite{Cardy:2022mhn}. From a complementary modular-invariance perspective,~\cite{Apolo:2023aho} studied the partition function of $T\bar T$-deformed CFTs on $\mathbb{T}^2$ and derived universal large-$c$ formulas for the density of states in both double- and single-trace deformations.

A treatment of local correlators was developed in~\cite{Cardy:2019qao}, where flow equations for correlation functions of local operators were derived, the derivation property of the deformation on the operator algebra was established, ultraviolet (UV) divergences were analyzed, and renormalized correlators together with a deformed operator product expansion and a Callan--Symanzik equation were obtained. In particular, for deformed CFTs on the plane, the momentum-space two-point function exhibits the characteristic $k$-dependent anomalous scaling implied by the $T\bar T$ flow. A complementary non-perturbative analysis based on JT gravity was carried out in~\cite{Aharony:2023dod}, where the same two-point function was computed and shown to display a distinct large-momentum behavior, reflecting the non-locality of the deformed theory at short distances. This analysis was later extended to $\mathbb{T}^2$ in~\cite{Barel:2024dgv}. In~\cite{Asrat:2020jsh,He:2020sun}, the flow equation and perturbative QFT methods were used to study the deformation of KdV charges and the corresponding modifications of the partition function. Later,~\cite{He:2022jyt} obtained the deformed CFT partition function on a genus-two Riemann surface $\Sigma^{(2)}$.

Subsequent work extended these results in several complementary directions. Perturbative analyses extended the study of deformed correlators to supersymmetric theories and to compact backgrounds: first-order correlation functions were derived for $2$d $\mathcal N=(1,1)$ and $\mathcal N=(2,2)$ SCFTs~\cite{He:2019vzf}, and the perturbative framework was generalized from $\mathbb{R}^2$ to $\mathbb{T}^2$~\cite{He:2020jsc}. From the random-geometry perspective~\cite{Cardy:2018sdv}, this framework was further developed to reproduce deformed correlators and to clarify its relation to the gravitational description~\cite{Hirano:2020ppu, Hirano:2025tkq}. More recently, the analysis has been studied by a systematic study of momentum-space two-point functions in unitary $T\bar T$-deformed CFTs~\cite{Aharony:2023dod}, as well as by higher-order treatments in general CFTs~\cite{He:2023qnh}. These developments suggest that the existence of deformed correlators is no longer the main focus; instead, their organization and kinematic characterization have become relevant.

More recently, a closely related deformation, known as root-$T\bar T$, has attracted considerable attention~\cite{Ferko:2022cix, Babaei-Aghbolagh:2022leo}. The operator can be defined as follows
\begin{equation}\label{rootTTbar Def}
O_{\sqrt{T\bar T}}
:=
\sqrt{\frac{1}{2}T^{\mu}{}_{\nu}T^{\nu}{}_{\mu}
-\frac{1}{4}\left(T^{\nu}{}_{\nu}\right)^2}\, .
\end{equation}
Correspondingly, the root-$T\bar T$ deformation of a QFT with undeformed action $S_0$ is defined by replacing the operator $O_{T\bar T}$ in Eq.~\eqref{TTbar Def} with $O_{\sqrt{T\bar T}}$, namely
\begin{equation}\label{rootTTbar Def}
\left\{
\begin{aligned}
S^{(0)} &= S_0, \\
\frac{\partial S^{(\gamma)}}{\partial \gamma}
&= \int_M \mathrm{d}^2x\, \sqrt{|g|}\, O_{\sqrt{T\bar T}}^{(\gamma)} \, ,
\end{aligned}
\right.
\end{equation}
where $\gamma$ is a dimensionless deformation parameter. At the classical level, the $T\bar T$ and root-$T\bar T$ flows commute, and the latter admits a geometric realization as a non-linear transformation of the dynamical variables~\cite{Ferko:2022cix}. At the quantum level, however, the structure of the deformation is much less understood. Existing work has mainly focused on the definition of the operator~\cite{Hadasz:2024pew} and on its gravitational formulation~\cite{Babaei-Aghbolagh:2024hti}, while a systematic analysis of local correlators is still lacking~\cite{Hadasz:2024pew}. From the field‑theoretic viewpoint, the root‑$T\bar{T}$ operator is non‑analytic and therefore does not fit directly into the standard perturbative framework developed for $T\bar{T}$ deformations.

In this paper, quasi-primary correlators in a two-dimensional CFT deformed simultaneously by $T\bar T$ and root-$T\bar T$ are studied within the random-geometry framework of~\cite{Hirano:2025tkq}. An effective geometric description of the combined flow is first constructed using a convenient parametrization of the frame field. In this framework, the deformed two-point function is obtained to all orders in the $T\bar T$ coupling and to leading order in the root-$T\bar T$ coupling. The leading correction to the three-point function under the combined deformation is then computed. Finally, the contributions of both $T\bar T$ and root-$T\bar T$ are shown to admit a reinterpretation as a weighted average of undeformed CFT correlators over conformal dimensions.

The paper is organized as follows. Sec.~\ref{set up} sets up the path integral formulation for the combined $T\bar T$ and root-$T\bar T$ deformation. Sec.~\ref{tworoot} contains the perturbative computation of the two-point function to leading order in the root-$T\bar T$ coupling and to all orders in $T\bar T$, and presents the kernels that express the deformed correlators as weighted averages over conformal dimensions. Sec.~\ref{threeroot} gives the leading perturbative correction to the three-point correlator. Conclusions and an outlook are provided in Sec.~\ref{conclusion}. Technical details are relegated to the appendices.

\section{Setup}\label{set up}
The combined $T\bar{T}$ and root-$T\bar{T}$ deformation is defined as the deformation that simultaneously satisfies the flow equations Eqs.~\eqref{TTbar Def} and~\eqref{rootTTbar Def}. In other words, for an undeformed theory $S_0$:
\begin{eqnarray}
S^{(0,0)} &=& S_0, \\
\frac{\partial S^{(\lambda,\gamma)}}{\partial \lambda} &=& -\int_M \mathrm{d}^2x \det(f^a_\mu) O_{T\bar{T}}^{(\lambda)} ,\\
\frac{\partial S^{(\lambda,\gamma)}}{\partial \gamma} &=& \int_M \mathrm{d}^2x \det(f^a_\mu) O_{\sqrt{T\bar{T}}}^{(\gamma)} .
\end{eqnarray}
The sign choices are made for convenience. Here, $f^a_\mu$ represents the fixed background frame of the deformed theory. As discussed in \cite{Babaei-Aghbolagh:2024hti}, introducing a dynamical frame $e^a_\mu$ makes the combined $T\bar{T}$ and root-$T\bar{T}$ deformation equivalent to coupling the original theory, at the saddle point of $e^a_\mu$, to the following gravitational theory~\cite{Babaei-Aghbolagh:2024hti}:
\begin{equation}\label{gravity}
S_{\text {grav}}\left[e, f\right]=\frac{1}{2 \lambda} \int \mathrm{d}^{2} x \operatorname{det} (e)\left(2+\ell_{1}^{2}-\ell_{2}-2 \ell_{1} \cosh \frac{\gamma}{2}+2 \sqrt{2 \ell_{2}-\ell_{1}^{2}} \sinh \frac{\gamma}{2}\right),
\end{equation}
where the Lorentz invariants $\ell_1 \equiv \mathrm{tr}(e^{-1}f)$ and $\ell_2 \equiv \mathrm{tr}[(e^{-1}f)^2]$ are introduced to express the action in a compact form. It is straightforward to verify that Eq.~\eqref{gravity} reduces to the dRGT action in the limit $\gamma \to 0$. The construction involves the dynamical metric tensor $g_{ij} = \delta_{ab} e^a_i e^b_j$. From a quantum gravity perspective, the deformed partition function is naturally defined as:
\begin{equation}\label{generalized partition function}
Z^{(\lambda,\gamma)}[\delta]:=\int \mathcal{D}g ~Z^{(0,0)}[g]~ e^{-S_{\text{grav}}[g]}.
\end{equation}
Here, $Z^{(0,0)}$ is the undeformed QFT partition function. Our primary interest in this work lies in the deformed correlation functions, which are defined, in terms of the partition function in Eq.~\eqref{generalized partition function}, as
\begin{equation}\label{generalized correlators}
\left\langle\prod_{A=1}^{n} \mathcal{O}\left(x_{A}\right)\right\rangle^{(\lambda,\gamma)}[\delta]:=\int \mathcal{D}g~ \left\langle\prod_{B=1}^{n} \mathcal{O}\left(x_{B}\right)\right\rangle^{(0,0)}[g]~e^{-S_{\text{grav}}[g]}.
\end{equation}
Henceforth, the case where the undeformed QFT is a CFT is considered, and quasi-primary fields $\mathcal{O}_{\Delta}$ with conformal dimension $\Delta = h + \bar h$ (with $h$ and $\bar h$ the holomorphic and antiholomorphic weights, respectively) are taken. The functional integration measure $\mathcal{D}g$ in the above expressions has not yet been properly defined. Let $\mathcal{M}$ be the configuration space of metric fields. To define the measure properly, an inner product of tensor on the cotangent space at each point $g \in \mathcal{M}$ is introduced as follows:
\begin{equation}\label{tensor inner product}
\left \langle h,h' \right \rangle_{\text{T}}:=\int \mathrm{d}^2x\sqrt{|g|}~ h_{ij} G^{ijkl} h_{kl},
\end{equation}
where $h_{ij}\equiv\delta g_{ij}$ is an infinitesimal one-form on the cotangent bundle $\text{T}^*\mathcal{M}$, and $G^{ijkl}$ is the well-known DeWitt supermetric~\cite{Dewitt}. This definition forces $G^{ijkl}$ to be symmetric under the exchange of the first two indices, the last two indices, and the interchange of the first pair with the second pair. To prevent the functional measure from introducing additional metric dynamics, $G^{ijkl}$ is required to be ultralocal, that is, at each point $x$ it depends only on the metric $g_{ij}(x)$ at that point and not on its derivatives. Under these requirements, $G^{ijkl}$ is determined up to an overall constant and can be written as:
\begin{equation}\label{supermetric}
G^{ijkl}=\frac{1}{2}(g^{ik}g^{jl}+g^{il}g^{jk}+C g^{ij}g^{kl}),
\end{equation}
where $C$ is an arbitrary constant~\footnote{To ensure the correct sign of the kinetic term of the supermetric, one typically requires $C>-1$.}. Consequently, the metric measure is defined by \cite{Dewitt,deWit1982,PhysRevD.43.1212,Mottola:1995sj}:
\begin{equation}
\int \mathcal{D}h \exp\left(-\frac{1}{2}\left \langle h,h \right \rangle_{\text{T}}\right):=1.
\end{equation}
Note that on a two-dimensional curved manifold, $h_{ij}\in T_g^*\mathcal{M}$ can always be decomposed into an infinitesimal diffeomorphism mode and a Weyl mode:
\begin{equation}\label{2d decompose}
h_{ij}=(L\alpha)_{ij}+(2\Phi+\nabla_\lambda\alpha^\lambda)g_{ij},
\end{equation}
where the covariant derivative $\nabla$ is compatible with $g_{ij}$, namely $\nabla_i g_{jk}=0$. The map $L$ is defined by
\begin{equation}
(L\alpha)_{ij}:=(\mathcal{L}_\alpha g)_{ij}-g_{ij}\nabla_\lambda\alpha^\lambda =\nabla_i\alpha_j+\nabla_j\alpha_i-g_{ij}\nabla_\lambda\alpha^\lambda.
\end{equation}
Here $\mathcal{L}_{\alpha}$ denotes the Lie derivative along the vector field $\alpha$, thus the map $L$ can be regarded as the operation of taking the traceless part of the Lie derivative. Eq.~\eqref{2d decompose} shows that the metric description can be naturally replaced by the parametrization in terms of $\alpha$ and $\Phi$~\cite{Polyakov:1981rd}. Similarly, the inner products for vector fields and scalar fields are defined respectively as \cite{Mottola:1995sj}:
\begin{eqnarray}
\left \langle \alpha,\alpha' \right \rangle_{\text{V}}&:=&\int \mathrm{d}^2x\sqrt{|g|} \alpha^i \alpha'_i~,\\
\left \langle \Phi,\Phi' \right \rangle_{\text{S}}&:=&\int \mathrm{d}^2x\sqrt{|g|} \Phi \Phi'.
\end{eqnarray}
Then the measures for $\alpha$ and $\Phi$ are defined by
\begin{eqnarray}
\int \mathcal{D}\alpha \exp\left(-\frac{1}{2}\left \langle \alpha,\alpha \right \rangle_{\text{V}}\right)&:=&1,\\
\int \mathcal{D}\Phi \exp\left(-\frac{1}{2}\left \langle \Phi,\Phi \right \rangle_{\text{S}}\right)&:=&1.
\end{eqnarray}
As will be seen later, using the dynamical variables $\alpha$ and $\Phi$ to perform the path integral allows for an analytic treatment~\cite{Polyakov:1981rd}. It is therefore necessary to replace $\mathcal{D}g$ in the definition by $\mathcal{D}\alpha\mathcal{D}\Phi$. This replacement, however, introduces an extra Jacobian factor $|J(g)|$. Using the definition~\eqref{tensor inner product} and the decomposition~\eqref{2d decompose}, one can determine $|J(g)|$ as follows:
\begin{equation}
\begin{aligned}\label{LdaggerL}
1&=|J(g)|\int\mathcal{D}\alpha \exp\left(-\frac{1}{2}\left \langle L\alpha,L\alpha\right \rangle_{\text{V}} \right)\int\mathcal{D}\sigma\exp(-8(1+C)\left \langle \sigma,\sigma \right \rangle_{\text{S}} )\\
&=\frac{|J(g)|}{\sqrt{-8(1+C)}}\text{det}^{-1/2}(L^\dagger L).
\end{aligned}
\end{equation}
In the first equality, a shift transformation $\sigma = \Phi + \frac{1}{2}\nabla_\lambda \alpha^\lambda$ is performed; this is a simple translation and therefore contributes a trivial Jacobian. The adjoint operator $L^\dagger$ is defined by \cite{PhysRevD.43.1212,Mottola:1995sj}
\begin{equation}
\left \langle \alpha,L^\dagger h \right \rangle_{\text{V}}:=\left \langle L\alpha,h \right \rangle_{\text{T}}.
\end{equation}
From this one can compute \cite{Mottola:1995sj}
\begin{equation}
(L^\dagger L)_{ij}=-2\delta_{ij}\nabla_\lambda\nabla^\lambda-\frac{R}{2}g_{ij},
\end{equation}
where $R$ is the scalar curvature associated with $g_{ij}$. Quantization essentially amounts to integrating over $\mathcal{M}$. This can be safely converted into an integral over the cotangent bundle $\text{T}^*\mathcal{M}$, because locally near $g\in\mathcal{M}$, a neighborhood of $g$ is diffeomorphic to $\text{T}_g^*\mathcal{M}$ via the exponential map. Hence \cite{Dewitt,deWit1982,Mottola:1995sj,Tolley:2019nmm}
\begin{equation}
\int\mathcal{D}h=\int \mathcal{D}g=\int \mathcal{D}\alpha\mathcal{D}\Phi |J(g)|.
\end{equation}
This completes the preparation of the measure. It remains to express the integrand of Eq.~\eqref{gravity} in terms of $\alpha$ and $\Phi$. A two-dimensional space is always locally conformally flat, which means
\begin{equation}\label{line element}
\mathrm{d}s^2_{\text{CFT}}=g_{ij}\mathrm{d}x^i\mathrm{d}x^j=e^{2\Phi(x)}\delta_{ij}\mathrm{d}y^i\mathrm{d}y^j.
\end{equation}
Here a flat coordinate system ${y^i}$ is defined such that $y^i = x^i + \alpha^i(x)$. The relation between correlation functions of a CFT on a curved two-dimensional space and those on flat space is~\cite{Polyakov:1981rd}
\begin{equation}\label{Liouville correlators}
\left\langle\prod_{A=1}^{n} \mathcal{O}\left(x_{A}\right)\right\rangle_{\text{CFT}}[g]=e^{-S_{\text{Liouville}}[g]}\left\langle\prod_{A=1}^{n} \mathcal{O}\left(y_{A}\right)\right\rangle_{\text{CFT}}[\delta],
\end{equation}
where $S_{\text{Liouville}}$ is called the Liouville action, explicitly given by
\begin{equation}\label{Liouville1}
S_{\text{Liouville}}[g]=\frac{c_0}{96\pi}\int \mathrm{d}^2x\sqrt{|g|}~R~ (\nabla^\mu\nabla_\mu)^{-1} R,
\end{equation}
where $c_0$ is the central charge of the CFT. Using the equality $R=-2e^{-2\Phi}\Box\Phi$, Eq.~\eqref{Liouville1} becomes \cite{Polyakov:1981rd,Mottola:1995sj}
\begin{equation}\label{Liouville2}
S_{\text{Liouville}}[g]=\frac{c_0}{24\pi}\int \mathrm{d}^2x~\Phi\Box\Phi,
\end{equation}
where $\Box\equiv\partial_i\partial^i$ is the two-dimensional Laplacian. Using Eq.~\eqref{line element}, the gravitational action~\eqref{gravity} can be factorized. There is, however, an inherent ambiguity: any frame field that satisfies Eq.~\eqref{line element} remains a solution under the action of an arbitrary orthogonal matrix. This is merely a parameterization redundancy—that is, a freedom in rewriting the metric in terms of the frame variables—rather than a gauge symmetry of the theory (the local Lorentz and diffeomorphism gauge symmetries of massive gravity have already been broken by fixing the reference metric). Nevertheless, this redundancy can be exploited to select a convenient explicit expression for the frame field, thereby yielding a tractable parameterized form of the gravitational action. Through a specific choice of frame $e_i^a$, the action~\eqref{gravity} can be parameterized as~\footnote{For computational convenience, the deformed QFT space is taken to be flat $\mathbb{R}^2$. For a curved background, correlation functions should be computed using the heat kernel expansion~\cite{DeWitt:1964mxt, Seeley:1967ea,Barvinsky:1985an,  Vassilevich:2003xt, Barvinskii:2024iqz, Barvinsky:2025bwu}; however, such an approach is expected to reliably capture only the UV regime.}
\begin{equation}\label{parameterized action}
S_{\text{grav}}(\alpha,\phi)=\frac{1}{\lambda}\int \mathrm{d}^2x\left(\frac{\phi^2}{s}+(s-\frac{1}{2}\partial_k\alpha^k)\sinh\frac{\gamma}{2}\sqrt{2\sigma_{ij}\sigma^{ij}}+\frac{1}{4}\left(3s-2\right)\alpha^i\Box\alpha_i \right).
\end{equation}
Here, the traceless symmetric tensor are defined by
\begin{equation}\sigma_{ij}\equiv\frac{1}{2}\left(\partial_i\alpha_j+\partial_j\alpha_i-\delta_{ij}\partial_k\alpha^k\right)
\end{equation}
to simplify the expression, and introduced the intrinsic Weyl mode
\begin{equation}\label{phi}
\phi=\Phi+\frac{1}{2}\ln\det\left(\frac{\partial y^i(\alpha)}{\partial x^j} \right)
+\frac{1}{2}s(\gamma)\sinh\frac{\gamma}{2}\sqrt{2\sigma_{ij}\sigma^{ij}}
\end{equation}
to decouple the infinitesimal diffeomorphism variable $\alpha^i$ from the total Weyl mode $\Phi$. In addition, the function $s(\gamma)\equiv\frac{1}{2-\cosh \frac{\gamma}{2}}$ is introduced. The calculation details and the specific parametrization choice for $e^a_i$ are presented in Appendix~\ref{appendixA}. As is apparent, the action consists of two parts: one arising from the root-$T\bar{T}$ contribution, and the other being a Gaussian kinetic term. In the limit $\gamma \to 0$, the Gaussian kinetic term exactly matches the result for pure $T\bar{T}$ deformation obtained in \cite{HiranoShigemori:2020,Hirano:2025tkq}.

Substituting Eqs.~\eqref{LdaggerL},~\eqref{Liouville correlators}, and~\eqref{parameterized action} into Eq.~\eqref{generalized correlators} yields \cite{RevModPhys.60.917,Mottola:1995sj}
\begin{equation}
\begin{aligned}
\left\langle\prod_{A=1}^{n} \mathcal{O}\left(x_{A}\right)\right\rangle^{(\lambda,\gamma)}[\delta]&=\mathcal{N}\int \mathcal{D}\alpha\mathcal{D}\Phi~ \left\langle\prod_{B=1}^{n} \mathcal{O}\left(y_{B}\right)\right\rangle_{\text{CFT}}[\delta]\\&\times\exp\bigg(-S_{\text{grav}}[\alpha,\Phi]-\frac{c}{24\pi}\int \mathrm{d}^2x~\Phi\Box\Phi\bigg),
\end{aligned}
\end{equation}
We will henceforth denote $\mathcal{N}$ as a normalization constant. Here
\begin{equation}
c=c_0-26+1.
\end{equation}
Here the factor $-26$ originates from the operator $L^\dagger L$, while the $+1$ comes from the Weyl anomaly of the scalar field $\Phi$~\cite{Polyakov:1981rd,RevModPhys.60.917,Mottola:1995sj}. After the Gaussian integration over $\Phi$ is performed, the expression in the exponent of the path integral becomes
\begin{equation}\label{Stotal}
S_{\text{tot}}(\alpha)=\frac{1}{\lambda}\int \text{d}^2x\left(s\sinh\frac{\gamma}{2}\sqrt{2\sigma_{ij}\sigma^{ij}}+\frac{1}{4}\left(3s-2-s\sinh^2\frac{\gamma}{2}\right)\alpha^i\Box\alpha_i\right).
\end{equation}
The coefficient $\lambda^{-2}$ is omitted here, as it contributes only a term proportional to $\lambda^2$ at subleading order (in the regularization sense) in the correlation function. The final expression for the correlation function can be written as
\begin{equation}\
\left\langle\prod_{A=1}^{n} \mathcal{O}\left(x_{A}\right)\right\rangle^{(\lambda,\gamma)}[\delta]=\mathcal{N}\int \mathcal{D}\alpha\mathcal{D}\Phi~ \left\langle\prod_{B=1}^{n} \mathcal{O}\left(y_{B}\right)\right\rangle_{\text{CFT}}[\delta]~e^{-S_{\text{tot}}[\alpha]}.
\end{equation}
The viewpoint of \cite{Hirano:2025tkq} is adopted here, and possible extra terms of the operator under conformal transformations are ignored, as they contribute only at subleading order.

\section{Perturbative two-point correlators of deformed CFTs}\label{tworoot}
\begin{figure}
    \centering
    \includegraphics[width=0.4\linewidth]{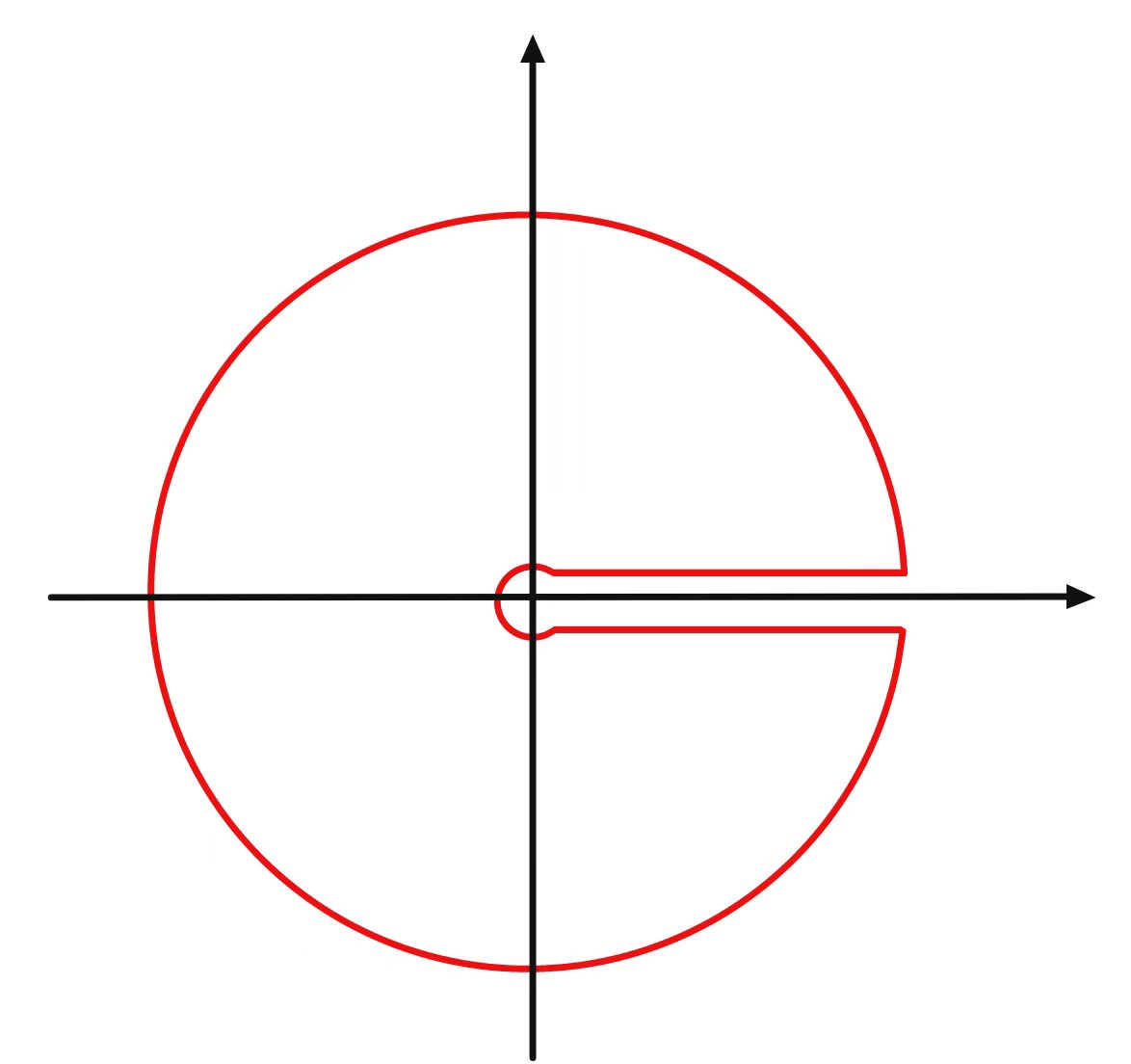}
    \caption{Schematic diagram of the contour encircling $\mathbb{R}_+$}
    \label{contour}
\end{figure}

To study the effect of the root-$T\bar{T}$ deformation at leading order, $\gamma$ is treated as a perturbative parameter and the total weight $e^{-S_{\mathrm{tot}}}$ is expanded in powers of $\gamma$. This yields
\begin{equation}
e^{-S_{\text{tot}}}
=e^{-S_0}\Bigg(1-\gamma S_1+\gamma^2\left(\frac{4S_1^2-S_0}{8}\right)+\gamma^3\left(\frac{3S_0S_1-4S_1-4S_1^3}{24}\right)+\mathcal{O}(\gamma^4)\Bigg),
\end{equation}
where
\begin{eqnarray}
S_{0}&\equiv&\frac{1}{4\lambda}\int \text{d}^2x~\alpha^i\Box\alpha_i,
\label{S0}\\
S_{1}&\equiv&\frac{1}{2\lambda}\int \text{d}^2x\sqrt{2\sigma_{ij}\sigma^{ij}}.
\end{eqnarray}
The CFT two-point correlation function of quasi-primary fields of conformal dimension $\Delta$ is
\begin{equation}\label{CFT twopoint}\left\langle O_\Delta(x_1)O_\Delta(x_2)\right\rangle_{\text{CFT}}=\frac{1}{\left|r\right|^{2\Delta}},\end{equation}
where $r\equiv x_1-x_2$. To first order in $\gamma$, the two-point correlation function of the deformed CFT is obtained as
\begin{equation}\label{first order of twopoint}
\left\langle \mathcal{O}_\Delta(x_1)\mathcal{O}_\Delta(x_2)\right\rangle_{\text{pert}}=-\frac{\gamma}{2\lambda}\int \mathcal{D}\alpha \frac{1}{\left | r+\alpha_{12}  \right |^{2\Delta} }\int \text{d}^2y \sqrt{2\sigma_{ij}(y)\sigma^{ij}(y)} e^{-S_0(\alpha) },
\end{equation}
where $\alpha_{12}\equiv\alpha(x_1)-\alpha(x_2)$. To handle the square root, the Schwinger parametrization~\cite{Hadasz:2024pew} is employed:
\begin{equation}\label{aba}
 \sqrt{2\sigma_{ij}(y)\sigma^{ij}(y)}=\frac{1}{4\sqrt{\pi}}\oint_{\mathbb{R}_+} \text{d}t~t^{-\frac{3}{2}}e^{-2t\sigma_{ij}(y)\sigma^{ij}(y)}.
\end{equation}
The validity of this integral is based on the analytic continuation of the Gamma‑function integral representation to $\mathbb{C}\setminus(\mathbb{Z}_+\cup{0})$. Note that the Gamma‑function parametrization is originally valid only for negative exponents, whereas the integral here is defined as a counterclockwise contour encircling the positive real axis, with the boundary consisting of the positive real axis starting from the origin, together with a circular contour at infinity in the complex plane (see Fig.~\ref{contour}). The expression in the exponent can be treated as an overall effective action. Namely, let
\begin{equation}
    S_{\text{eff}}(\alpha,y,t)=2t\sigma_{ij}(y)\sigma^{ij}(y)+\frac{1}{4\lambda}\int \text{d}^2x~\alpha^i\Box\alpha_i. 
\end{equation}
The presence of the first term introduces a crucial nontrivial structure: a kinetic term localized at the point $y$ without spatial integration. Physically, this can be interpreted as introducing a specific point-like defect into the background $\mathbb{R}^2$.~The usual approach is to forcibly introduce generalized functions to turn it into a local Lagrangian density,~i.e.
\begin{equation}
    S_{\text{eff}}(\alpha,y,t)=2t\int \text{d}^2x ~\delta^2(y-x)\sigma_{ij}(x)\sigma^{ij}(x)+\frac{1}{4\lambda}\int \text{d}^2x~\alpha^i\Box\alpha_i. 
\end{equation}
Here,~$\delta^2(x-y)$ is the two-dimensional Dirac function. Through integration by parts and ignoring boundary terms, a direct calculation shows that
\begin{equation}\label{Seff}
    S_{\text{eff}}(\alpha,y,t)=\int \text{d}^2x~\alpha^i(x)\hat{O}_{ij}\alpha^j(x).
\end{equation}
Which means the effective action can be expressed in a quadratic kinetic form,~where
\begin{eqnarray}
\hat{O}_{ij}&\equiv&\frac{1}{4\lambda}\delta_{ij}\Box-2t\delta_{ij}\delta^2(x-y)\Box-2t\hat{M}_{ij},\\
\hat{M}_{ij}&\equiv&\left (\partial_j\delta^2(y-x)\right)\partial_i-\left (\partial_i\delta^2(y-x)\right)\partial_j+\delta_{ij}\left (\partial^k\delta^2(y-x)\right)\partial_k.
\end{eqnarray}
The Green's function of $\hat{O}_{ij}$ is~\footnote{Admittedly, from a rigorous mathematical perspective, directly performing the path integral for the non‑elliptic operator $\hat O{ij}$ requires careful justification. A more systematic framework based on distribution theory would be needed, but this lies beyond the perturbative scope of the present work. Nevertheless, under the point‑splitting regularization employed here, the non‑trivial functional structure of the correlation functions is retained. The current treatment is therefore adopted as a reasonable technical tool, and further investigation of this issue within a more rigorous mathematical framework is left for future work.}
\begin{equation}\label{Green function}
G_{ij}(x,x',y,t)=\frac{2\lambda}{\pi}\bigg(S(x,x',y)\delta_{ij}+A(x,x',y)\varepsilon_{ij}\bigg),   
\end{equation}
with
\begin{eqnarray}
S(x,x',y,t)&\equiv&\ln\left(\frac{\left|x-x'\right|}{\varepsilon}\right)-\frac{4\lambda t}{\pi}\frac{(x'-y)\cdot(x-y)}{\left|x'-y\right|^2\left|x-y\right|^2},\label{symmetric part}\\
A(x,x',y,t)&\equiv&-\frac{4\lambda t}{\pi}\frac{\bigg((x'-y)\times(x-y)\bigg)_3}{\left|x'-y\right|^2\left|x-y\right|^2}.\label{anti-symmetric part} 
\end{eqnarray}
Here, $x\cdot y\equiv x_iy^i$ and $(x\times y)_3\equiv x_1y_2-x_2y_1=\varepsilon^{ij}x_iy_j$~\footnote{The convention adopted for the Levi-Civita symbols is $\varepsilon^{12}=\varepsilon_{12}=+1$.}. The calculation process is listed in Appendix~\ref{appendixB}. To convert the $\alpha$-dependence in the CFT two-point function~\eqref{CFT twopoint} from a complicated denominator into an exponential form, which can then be combined with the quadratic term in $S_{\mathrm{eff}}(\alpha, y, t)$, a Fourier transformation to momentum space is employed. This follows the strategy of~\cite{Hirano:2025tkq}, where the power-law form of the CFT two-point function is expressed as a Gaussian integral in momentum space, thereby rendering the overall path integral a tractable Gaussian integral. Hence, in momentum space:
\begin{equation}
\begin{aligned}\label{momentum decompose}
|r+\alpha_{12}|^{-2\Delta}&=\frac{1}{\Gamma(\Delta) }\int_0^\infty \text{d}\left(\frac{1}{4u}\right)~\left(\frac{1}{4u}\right)^{\Delta-1}e^{-\frac{ |r+\alpha_{12}|^2}{4u}}\\
&=\frac{2^{-2\Delta}}{\pi\Gamma(\Delta) }\int_0^\infty \text{d}u~u^{-\Delta}\int_{\mathbb{R}^2}\text{d}^2k~e^{ik\cdot(r+\alpha_{12})-u|k|^2}\\
&=\frac{2^{-2\Delta}}{\pi }\frac{\Gamma(1-\Delta)}{\Gamma(\Delta)}\int_{\mathbb{R}^2}\text{d}^2k~e^{ik\cdot(r+\alpha_{12})}|k|^{2(\Delta-1)}.
\end{aligned}
\end{equation}
With the Gamma functions understood via analytic continuation, the expression remains valid for arbitrary real $\Delta\in\mathbb{R}$, and the undeformed CFT two-point function in Eq.~\eqref{first order of twopoint} can thus be decomposed in momentum space using Eq.~\eqref{momentum decompose}:
\begin{equation}\label{calculation1}
\begin{aligned}
    \left\langle \mathcal{O}_\Delta(x_1)\mathcal{O}_\Delta(x_2)\right\rangle_{\text{pert}}&=-\frac{\gamma}{2\lambda}\frac{1}{4\sqrt{\pi}}\frac{\Gamma(1-\Delta)}{\Gamma(\Delta)}\frac{1}{2^{2\Delta}\pi }\\&\times\int \mathcal{D}\alpha \int_{\mathbb{R}^2} \text{d}^2y\int_{\mathbb{R}^2} \text{d}^2k~e^{ik\cdot(r+\alpha_{12})}|k|^{2\Delta-2}\oint_{\mathbb{R}_+} \text{d}t~t^{-\frac{3}{2}} e^{-S_{\text{eff}}(\alpha,y,t)}.
    \end{aligned}
\end{equation}
Since the $\alpha$-dependence in Eq.~\eqref{calculation1} is a standard Gaussian integral with source term $e^{ik\cdot\alpha_{12}}$, and the operator $\hat{O}_{ij}$ is self-adjoint (see Appendix~\ref{appendixC}), the Gaussian integration can be performed to obtain
\begin{equation}\label{calculation2}
\begin{aligned}
    \left\langle \mathcal{O}_\Delta(x_1)\mathcal{O}_\Delta(x_2)\right\rangle_{\text{pert}}&=-\frac{\gamma}{2\lambda}\frac{1}{4\sqrt{\pi}}\frac{\Gamma(1-\Delta)}{\Gamma(\Delta)}\frac{1}{2^{2\Delta}\pi }\\&\times \int \text{d}^2y\oint_{\mathbb{R}_+} \text{d}t~t^{-\frac{3}{2}}Z_{\text{eff}}(t)\int_{\mathbb{R}^2} \text{d}^2k~e^{ik\cdot r}|k|^{2\Delta-2} e^{\frac{\lambda}{\pi}k^2S(x_1,x_2,y,t)}.
\end{aligned}
\end{equation}
Here, $S(x_1,x_2,y,t)$ is the symmetric part~\eqref{symmetric part} of the Green's function and the partition function of the effective action is defined as $Z_{\mathrm{eff}}(t)\equiv\int \mathcal{D}\alpha~ e^{-S_{\mathrm{eff}}(\alpha,y,t)}$, which is independent of the spatial point $y$ because both the action $S_{\mathrm{eff}}(\alpha,y,t)$ and the integration measure $\mathcal{D}\alpha$ are translation invariant. Expanding $e^{\frac{\lambda}{\pi}k^2S(x_1,x_2,y,t)}$ as a power series and integrating over $k$, one obtains~\footnote{This follows from $$\frac{\Gamma(1-\Delta)}{\Gamma(\Delta)}\frac{\Gamma(\Delta+n)}{\Gamma(1-\Delta-n)}=(-1)^n\left(\frac{\Gamma(\Delta+n)}{\Gamma(\Delta)}\right)^2.$$}
\begin{equation}\label{calculation3}
\begin{aligned}
   \left\langle \mathcal{O}_\Delta(x_1)\mathcal{O}_\Delta(x_2)\right\rangle_{\text{pert}}
   &=-\frac{\gamma}{2\lambda}\frac{1}{4\sqrt{\pi}}\sum_{n=0}^\infty\frac{(-\frac{4\lambda}{\pi})^n}{n!}\bigg(\frac{\Gamma(\Delta+n)}{\Gamma(\Delta)}\bigg)^2 \frac{1}{|r|^{2\Delta+2n}}\\&\times \oint_{\mathbb{R}_+} \text{d}t~t^{-\frac{3}{2}}Z_{\text{eff}}(t)\int \text{d}^2y~S^n(x_1,x_2,y,t).
\end{aligned}
\end{equation}
The integral over $y$ in Eq.~\eqref{calculation2} is evaluated as follows:
\begin{equation}
\int \text{d}^2y~S^n(x_1,x_2,y,t)=\sum_{p=0}^n\binom{n}{p}\left(-\frac{4\lambda t}{\pi} \right)^p\ln^{n-p}\bigg(\frac{|r|}{\varepsilon } \bigg)\int \text{d}^2y \left(\frac{(x_1-y)\cdot(x_2-y)}{|x_1-y|^2|x_2-y|^2}\right)^p ,
\end{equation}
For $p=0$, the integral is non-renormalizable. For $p\ge 1$, the integral has a standard closed form, whose finite part can be extracted via dimensional regularization. The result reads
\begin{equation}
\begin{aligned}
&~~~~~\int \text{d}^Dy \left(\frac{(x_1-y)\cdot(x_2-y)}{|x_1-y|^2|x_2-y|^2}\right)^p\\&=2^{-p}\pi^{\frac{D}{2}}|r|^{D-2p}\sum_{k=0}^{p} \sum_{j=0}^{p-k}\binom{p}{k}\binom{p-k}{j}(-1)^{p-k-j} \frac{\Gamma(2 p-j-k-\frac{D}{2} ) \Gamma(\frac{D}{2}-p+j) \Gamma(\frac{D}{2}-p+k)}{\Gamma(p-j) \Gamma(p-k) \Gamma(D-2 p+j+k)} ,\\
&\xlongequal{D\longrightarrow2}\pi|r|^{2-2p}\left(A_p\ln\left(\frac{|r|}{\varepsilon}\right)+B_p\right).
\end{aligned}
\end{equation}
The expressions for $A_p$ and $B_p$ are
\begin{eqnarray}
A_p&=&2^{-p} \sum_{k=0}^{p} \sum_{j=0}^{p-k} \binom{p}{k} \binom{p-k}{j} (-1)^{p-k-j} \, \text{Res}_{p,j,k},\\
B_p&=&2^{-p} \sum_{k=0}^{p} \sum_{j=0}^{p-k} \binom{p}{k} \binom{p-k}{j} (-1)^{p-k-j} \left(\text{Const}_{p,j,k}+\frac{\text{Res}_{p,j,k}}{2}\ln\pi\right),
\end{eqnarray}
where $\gamma_E$ is the Euler–Mascheroni constant,~$\text{Res}_{p,j,k}$ and $\text{Const}_{p,j,k}$ are the coefficient corresponding to the $\delta^{-1}$ and $\delta^{0}$ term of the function
\begin{equation}
g_{p, j, k}(\delta)=\frac{\Gamma(2 p-j-k-1-\frac{\delta}{2}) \Gamma(1-p+j+\frac{\delta}{2}) \Gamma(1-p+k+\frac{\delta}{2})}{\Gamma(p-j) \Gamma(p-k) \Gamma(2-2 p+j+k+\delta)}   
\end{equation}
respectively. The integral over $t$ is evaluated as follows:
\begin{equation}\label{cp}    C_p\equiv\oint_{\mathbb{R}_+} \text{d}t~t^{-\frac{3}{2}+p}Z_{\text{eff}}(t)=\mathcal{N}~\Gamma(p-\frac{1}{2})\left \langle (2\sigma_{ij}\sigma^{ij})^{\frac{1}{2}-p } \right \rangle_0.
\end{equation}
Here $\langle\cdots\rangle_0$ denotes the expectation value with respect to $S_0$. Since this expectation value is translation invariant, the expression in Eq.~\eqref{cp} can be evaluated at any point in $\mathbb{R}^2$; therefore, $C_p$ itself is a nonlocal c-number that acts as a constant factor in the correlation functions. A global change of variables gives $C_{p+1}/C_p \sim \lambda^{-1}$, so that the dimension difference between adjacent coefficients in $\{C_p, p\ge 1\}$ is $[\text{length}]^2$. Consequently, the root-$T\bar{T}$ contribution to the two-point function appears as a perturbative correction:
\begin{equation}\label{combined twopoint}
    \left\langle \mathcal{O}_{\Delta}\left(x_{1}\right) \mathcal{O}_{\Delta}\left(x_{2}\right)\right\rangle_{\text {pert }}= \sum_{m=0}^{\infty}\sum_{p=1}^{\infty}\frac{\ln ^{m}\bigg(\frac{|r|}{\varepsilon}\bigg)}{|r|^{2 \Delta+2 m+4 p-2}}\left[\mathcal{A}_{p, m}\ln\bigg(\frac{|r|}{\varepsilon}\bigg)+\mathcal{B}_{p, m}\right],
\end{equation}
where 
\begin{eqnarray}
\mathcal{A}_{p, m} & =&-\frac{\gamma \sqrt{\pi}}{8 \lambda}  \frac{(-1)^m}{m!p!}\left(\frac{4 \lambda}{\pi}\right)^{m+2p}\left(\frac{\Gamma(\Delta+m+p)}{\Gamma(\Delta)}\right)^{2}C_{p} A_{p}, \\
\mathcal{B}_{p, m} & =&-\frac{\gamma \sqrt{\pi}}{8 \lambda}  \frac{(-1)^m}{m!p!}\left(\frac{4 \lambda}{\pi}\right)^{m+2p}\left(\frac{\Gamma(\Delta+m+p)}{\Gamma(\Delta)}\right)^{2} C_{p} B_{p}.
\end{eqnarray}
As expected, this result shares the same logarithmic and power-law structure as the pure $T\bar{T}$ two-point function~\cite{Cardy:2019qao}. However, here a double sum appears: the index $m$ originates from the series expansion of the pure $T\bar{T}$ part, while the index $p$ comes from the Schwinger parametrization of the square root in the root-$T\bar{T}$ term. This reflects the mixing effect of the two deformations at the perturbative level.

Moreover, the two-point correlation function of the deformed CFT in the pure $T\bar{T}$ case ($\gamma \to 0$) has been computed in~\cite{Hirano:2025tkq} and is given by:
\begin{equation}\label{twopoint}
\langle |r+\alpha_{12}|^{-2\Delta} \rangle_0=\sum_{n=0}^{\infty}(-1)^{n} \frac{1}{n!}\left(\frac{4 \lambda}{\pi}\right)^{n}\left(\frac{\Gamma(\Delta+n)}{\Gamma(\Delta)}\right)^2\frac{\ln ^{n}\left(\frac{\left|r\right|}{\varepsilon}\right)}{\left|r\right|^{2(\Delta+n)}}.
\end{equation}
This result matche that of ~\cite{Cardy:2019qao}, demonstratings the validity of this framework. Next, we make a further analysis of Eq.~\eqref{twopoint}. Note that it can be rewritten as:
\begin{equation}\label{operator of twopoint}
\left\langle \mathcal{O}_{\Delta}\left(x_{1}\right) \mathcal{O}_{\Delta}\left(x_{2}\right)\right\rangle_{T\bar T}=\left(\sum_{n=0}^{\infty}\frac{1}{n!}\left(\frac{2 \lambda }{\pi}\right)^{n}\left(\frac{\Gamma(\Delta+n)}{\Gamma(\Delta)}\right)^2 e^{n\partial_\Delta}\partial^n_\Delta\right)\left\langle \mathcal{O}_{\Delta}\left(x_{1}\right) \mathcal{O}_{\Delta}\left(x_{2}\right)\right\rangle_{\text{CFT}}.
\end{equation}
The operator on the right-hand side of Eq.~\eqref{operator of twopoint} has no explicit dependence on $x_1$ or $x_2$. Hence, the $T\bar{T}$ deformation of the two-point function is equivalent to the action of a single operator on the undeformed two-point function. By the Schwartz kernel theorem, this operator possesses a two‑variable distribution kernel $K_{T\bar{T}}(\Delta;\Delta')$, which allows the deformation to be expressed as a generalized integral:
\begin{equation}\label{kernel}
K_{T\bar T}(\Delta, \Delta') = \int_{-\infty}^{\infty} \frac{\text{d}k}{2\pi} \, e^{ik(\Delta - \Delta')} \, {}_2F_1\!\left( \Delta, \Delta; 1; \frac{2\lambda}{\pi} i k e^{ik} \right).
\end{equation}
Consequently, the $T\bar T$ deformed two-point correlator can be rewritten as
\begin{equation}\label{kerneleq}
\left\langle \mathcal{O}_{\Delta}\left(x_{1}\right) \mathcal{O}_{\Delta}\left(x_{2}\right)\right\rangle_{T\bar T}=\int \text{d}\Delta' K_{T\bar T}(\Delta;\Delta')\left\langle \mathcal{O}_{\Delta'}\left(x_{1}\right) \mathcal{O}_{\Delta'}\left(x_{2}\right)\right\rangle_{\text{CFT}}.
\end{equation}
Further details are given in Appendix~\ref{appendixD}. Eq.~\eqref{kerneleq} shows that the two-point function of the $T\bar T$-deformed CFT can be obtained as a weighted average of two-point functions of CFTs with all possible conformal dimensions $\Delta'$, with the kernel $K_{T\bar T}(\Delta;\Delta')$ serving as the weight function. Similarly, the kernel of Eq.~\eqref{combined twopoint} can be computed as
\begin{equation}\label{combined kernel}
\begin{aligned}
K_{\mathrm{comb}}\left(\Delta, \Delta^{\prime}\right)&=-\frac{\gamma \sqrt{\pi}}{8 \lambda} \sum_{p=1}^{\infty} \frac{C_{p}}{p!}\left(\frac{4 \lambda}{\pi}\right)^{2 p}\left(\frac{\Gamma(\Delta+p)}{\Gamma(\Delta)}\right)^2\\&\times\left(B_{p}+\frac{A_{p}}{2} \frac{\partial}{\partial \Delta^{\prime}}\right) K_{T \bar{T}}\left(\Delta+p ; \Delta^{\prime}-p+1\right) .
\end{aligned}
\end{equation}
Eq.~\eqref{combined kernel} shows that the kernel for the combined deformation is obtained from the pure $T\bar T$ kernel by a weighted sum over the root‑$T\bar T$ expansion index $p$, with each term involving a shift of conformal dimensions and a derivative with respect to $\Delta'$. This structure reflects the interplay of the two deformations in the weighted‑average representation.

\section{Perturbative three-point correlators in deformed CFTs}\label{threeroot}
The standard CFT three-point correlation function is given by
\begin{equation}
\left\langle \mathcal{O}_{\Delta_1}(x_1)\mathcal{O}_{\Delta_2}(x_2)\mathcal{O}_{\Delta_3}(x_3)\right\rangle_{\mathrm{CFT}} = \prod_{\mathrm{cycl}} \frac{1}{|x_A - x_B|^{2\delta_C}}.
\end{equation}
Here the notation $\delta_C \equiv \frac{1}{2}(\Delta_A + \Delta_B - \Delta_C)$ is introduced for distinct $A, B, C$, and $\mathrm{cycl}$ indicates that $A, B, C$ are assigned to the even permutations of $1, 2, 3$, respectively. Transforming each CFT factor into momentum space via the identity~\eqref{momentum decompose} yields the deformed three-point correlator at first order in $\gamma$:
\begin{equation}
\begin{aligned}
&\left\langle \mathcal{O}_{\Delta_1}(x_1)\mathcal{O}_{\Delta_2}(x_2)\mathcal{O}_{\Delta_3}(x_3)\right\rangle_{\text{pert}}=-\frac{\gamma}{2\lambda}\frac{1}{4\sqrt{\pi}}\int \text{d}^2y\oint_{\mathbb{R}_+} \text{d}t~t^{-\frac{3}{2}}~~~~~\\
&\times\prod_{\text{cycl}} \frac{\Gamma\left(1-\delta_{A}\right)}{\pi 2^{2 \delta_{A} }\Gamma\left(\delta_{A}\right)} \int_{\mathbb{R}^2} \text{d}^{2} k_{A} \left|k_{A}\right|^{2\left(\delta_{A}-1\right)}\int\mathcal{D}\alpha ~e^{-S_{\text{eff}}+i k_{A} \cdot\left(r_{B C}+\alpha_{B C}\right)}.
\end{aligned}
\end{equation}
Here the notations $r_{AB}\equiv x_A-x_B$ and $\alpha_{AB}\equiv \alpha(x_A)-\alpha(x_B)$ are introduced. Taking the external source $J^i(x)=i\sum_{\mathrm{cycl}}k^i_A\big(\delta^2(x-x_B)-\delta^2(x-x_C)\big)$ and performing the integral over $\alpha$ gives~\footnote{For notational simplicity, the $y$ and $t$ dependence is suppressed.}:
\begin{equation}\label{calculation4}
\begin{aligned}
&\left\langle\mathcal{O} _{\Delta_1}(x_1)\mathcal{O}_{\Delta_2}(x_2)\mathcal{O}_{\Delta_3}(x_3)\right\rangle_{\text{pert}}=-\frac{\gamma}{2\lambda}\frac{1}{4\sqrt{\pi}}\int \text{d}^2y\oint_{\mathbb{R}_+} \text{d}t~t^{-\frac{3}{2}}Z_{\text{eff}}(t)\\
&\times\prod_{\text{cycl}} \frac{\Gamma\left(-\delta_{A}+1\right)}{\pi 2^{2 \delta_{A} }\Gamma\left(\delta_{A}\right)} \int_{\mathbb{R}^2} \text{d}^{2} k_{A} e^{i k_{A} \cdot r_{B C}}\left|k_{A}\right|^{2\left(\delta_{A}-1\right)}\exp\left(\frac{\lambda}{\pi}\mathcal{S}(x_A,x_B,x_C)\right),
\end{aligned}
\end{equation}
where
\begin{eqnarray}
\mathcal{S}(x_A,x_B,x_C)&\equiv|k_A|^2S(x_B,x_C)-k_A\cdot k_B\tilde S(x_A,x_B,x_C)-(k_A\times k_B)_3\tilde A(x_A,x_B,x_C), \\
\tilde{S}(x_A,x_B,x_C)&\equiv S(x_B,x_C)+S(x_C,x_A)-S(x_B,x_A),~~~~~~~~~~~~~~~~~~~~~~~~~~~~~~~~~~~~~~~\\
\tilde{A}(x_A,x_B,x_C)&\equiv A(x_B,x_C)+A(x_C,x_A)-A(x_B,x_A).~~~~~~~~~~~~~~~~~~~~~~~~~~~~~~~~~~~~~~~
\end{eqnarray}
Here, $S(x,x')$ and $A(x,x')$ are the symmetric part~\eqref{symmetric part} and antisymmetric part~\eqref{anti-symmetric part} of the Green's function, respectively. A Taylor expansion of this integral produces lengthy polynomial expressions that offer no additional physical insight. For this reason, the explicit expanded form is not presented here. The analytic structure of the three‑point correlation function is therefore examined only at leading order in the $T\bar T$ deformation parameter $\lambda$:
\begin{equation}
\begin{aligned}\label{calculation5}
\int \text{d}^2y \Bigg(1+\sum_{\text{cycl}}\frac{\lambda}{\pi}\mathcal{S}(x_A,x_B,x_C) \Bigg)
&\xlongequal{\text{reg.}}\frac{4\lambda^2t}{\pi}\sum_{\text{cycl}}\Bigg(|k_A|^2\left(2\ln\left(\frac{r_{BC}}{\varepsilon}\right)+(\gamma_E+\ln\pi)\right)\\&+k_A\cdot k_B\left(2\ln\left(\frac{\varepsilon|r_{AB}|}{|r_{AC}||r_{CB}|}\right)-(\gamma_E+\ln\pi)\right)\Bigg).
\end{aligned}
\end{equation}
Here $\tilde A(x_A,x_B,x_C,y,t)$ does not contribute to the integral because the corresponding angular integration vanishes by rotational invariance.
\begin{equation}
\left(r_{AB}\times\int_{\mathbb{R}^2} \text{d}^2y~\frac{y}{|y|^2|r_{BA}-y|^2}\right)_3\propto (r_{AB}\times r_{AB})_3=0.
\end{equation}
Physically, this cancellation indicates that the antisymmetric kernel, which originates from the $\varepsilon_{ij}$ part of the Green's function, does not contribute to the three-point correlator at leading order. Substituting Eq.~\eqref{calculation5} into the three-point correlator yields:
\begin{equation}\label{combined threepoint}
\left\langle\mathcal{O} _{\Delta_1}(x_1)\mathcal{O}_{\Delta_2}(x_2)\mathcal{O}_{\Delta_3}(x_3)\right\rangle_{\text{leading}}=-\frac{2\gamma\lambda}{\pi^{\frac{3}{2}}}C_1\frac{\sum_{\text{cycl}}\frac{\delta_A^2}{|r_{BC}|^2}\left(2\ln\frac{|r_{BC}|}{\varepsilon}+(\gamma_E+\ln\pi)\right)}{|r_{12}|^{2\delta_3}|r_{23}|^{2\delta_1}|r_{13}|^{2\delta_2}}.
\end{equation}
Here, $C_1$ is defined in Eq.~\eqref{cp} with $p=1$. This is the leading-order contribution in the $T\bar T$ parameter to the three-point correlation function of the deformed CFT, with the root-$T\bar T$ deformation treated as a perturbation. Eq.~\eqref{combined threepoint} shows that, when the combined deformation is treated as a perturbation, logarithmic corrections appear in the three-point correlator. The logarithmic terms arise from UV divergences and depend on the renormalization scheme, while the power-law factors dominate the IR behavior. This mixing between the two types of terms reflects that the root-$T\bar{T}$ deformation affects the theory both in the UV (via the square-root operator) and in the IR (through induced anomalous dimensions). Such behavior is typical of nonlocal irrelevant deformations, which break scale invariance yet remain analytically tractable. Furthermore, the cyclic sum over indices indicates that the correction is not simply a product of individual two-point anomalies, but a genuine three-point mixing effect.

\section{Conclusions and outlook}\label{conclusion}

Quasi-primary correlators in two-dimensional CFTs deformed simultaneously by $T\bar{T}$ and root-$T\bar{T}$ are studied within a path-integral framework based on the geometric realization of deformations~\cite{Babaei-Aghbolagh:2024hti}. Starting from the gravitational description of the combined deformation, the problem was reformulated in terms of fluctuating geometric variables, and local correlation functions were analyzed perturbatively in the root-$T\bar T$ coupling while retaining the full $T\bar T$ dependence.

The main results are as follows. The deformed two-point function was obtained to all orders in the $T\bar T$ coupling and to leading order in the root-$T\bar T$ coupling, and the leading perturbative correction to the three-point function in the combined deformation was derived. The pure $T\bar T$ two-point function was shown to admit a kernel representation as a weighted average of undeformed CFT correlators over conformal dimensions, and the corresponding kernel for the combined deformation was constructed. In this picture, the deformation reorganizes the undeformed CFT data through a geometric averaging procedure.

The Schwinger parametrization provides a perturbative treatment of the square root but does not resum the full root‑$T\bar{T}$ contribution. In the present work, this term is treated only at leading order. Whether the square root can be resummed to all orders while keeping the path integral measure of Sec.~\ref{set up} unchanged remains to be clarified.


Several directions deserve further study. It would be natural to extend the present analysis to higher orders in the root-$T\bar T$ coupling and to higher-point correlators, and to generalize the construction to compact or curved backgrounds. Another important question concerns holography. While progress has been made in the AdS$_3$/CFT$_2$ description of root-$T\bar T$ deformations at the level of boundary conditions and partition functions~\cite{Ebert:2023tih, He:2025fdz}, The holographic interpretation of local correlators is not yet fully characterized. It would be interesting to investigate whether the correlators obtained here admit a direct bulk description and whether the weighted-average picture identified in this work has a natural counterpart in AdS$_3$ gravity.

\begin{acknowledgments}
We thank H. Babaei-Aghbolagh, Bin Chen, Shinji Hirano, Jue Hou, Yun-Ze Li, Yue-Ming Ma, Yun-Fei Xie, and Yang Yu for helpful discussions. 
This work was supported by the National Natural Science Foundation of China (Grants No. 12475056,
No. 12475053, No. 12588101, No. 12235016, and No. 12247101), Gansu Province's Top Leading Talent Support Plan, the
Fundamental Research Funds for the Central Universities (Grant No. lzujbky-2025-jdzx07),  the Natural Science Foundation of Gansu Province (No. 22JR5RA389 and No. 25JRRA799), and the 111 Project (Grant No. B20063).
\end{acknowledgments}

\appendix
\section{Derivation of Eq.~\eqref{parameterized action}}\label{appendixA}
For simplicity, the deformed QFT is taken to live on flat $\mathbb{R}^2$ with $f_i^a = \delta_i^a$, so that $\ell_1 = \operatorname{tr}(e^{-1})$ and $\ell_2 = \operatorname{tr}[(e^{-1})^2]$. To simplify the gravitational action~\eqref{gravity} in terms of the frame field, it is noted that in two dimensions any $2\times2$ matrix $A$ satisfies the following property:
\begin{equation}\label{A property}
\text{det}A=\frac{1}{2}\left((\text{tr}A)^2-(\text{tr}A^2)\right)=\frac{\text{tr}A}{\text{tr}(A^{-1})}.
\end{equation}
Using this property,~one can rewrite the action \eqref{gravity} as:
\begin{equation}\label{action2}
    S_{\text{grav}}[e]=\frac{1}{\lambda}\int \text{d}^2x\left(\text{det}(e)+1-\cosh\frac{\gamma}{2}\text{tr}(e)+\sinh\frac{\gamma}{2}\sqrt{(\text{tr}(e))^2-4\text{det}e}\right).
    \end{equation}
Starting from Eq.~\eqref{line element}, a solution for the frame field $e_i^a$ is obtained by iteratively taking the square root of the metric. To this end, $e_i^a$ is expanded as
\begin{equation}\label{D}
e^a_i=e^\Phi\left(\delta^a_{~i}+D^{a(1)}_{~i}+D^{a(2)}_{~i}+D^{a(3)}_{~i}+...\right),
\end{equation}
where $D^{(n)a}_{~~~~i}$ are the same order as $\alpha^m\phi^{n-m}$, with $1\le m \le n$, $m\in\mathbb{Z}$. Substituting Eq.~\eqref{D} into Eq.~\eqref{line element} and matching terms at each order yields~\footnote{The symmetrized and antisymmetrized brackets are defined as $A_{(ij)}=\frac{1}{2}(A_{ij}+A_{ji})$ and $A_{[ij]}=\frac{1}{2}(A_{ij}-A_{ji})$, respectively.}
\begin{equation}\label{each order}
     \left\{\begin{matrix}
D^{(1)}_{(ij)}&=&\partial_{(i}\alpha_{j)} \\
2D^{(2)}_{(ij)}+D^{k(1)}_{~i}D^{(1)}_{kj}&=&\partial_i\alpha_k\partial_j\alpha^k\\
2D^{(3)}_{(ij)}+D^{k(1)}_{~i}D^{(2)}_{kj}+D^{k(2)}_{~i}D^{(1)}_{kj}&=&0 \\
...&
\end{matrix}\right.
\end{equation}
From the first equation in Eq.~\eqref{each order}, it follows that $D^{(1)}_{ij}$ differs from $\partial_i\alpha_j$ by an arbitrary antisymmetric tensor, denoted as $\chi\varepsilon_{ij}$. That is:
\begin{equation}\label{D1}
D^{(1)}_{ij}=\partial_i\alpha_j+\chi\varepsilon_{ij}.
\end{equation}
As discussed in Sec.~\ref{set up}, an appropriate choice of $\chi$ is sought to optimize the gravitational action as much as possible. Substituting Eq.~\eqref{D1} into Eq.~\eqref{D} determines $D^{(2)}_{ij}$ (up to an antisymmetric term). Proceeding iteratively, $D^{(n)}_{ij}$ can similarly be determined (also up to an antisymmetric term). Among these additional antisymmetric terms, the antisymmetric parts of $D^{(2)}$ and higher orders do not contribute; only $\chi$ yields actual contributions to the second order. Consequently, all antisymmetric terms except $\chi$ are henceforth set to zero. Following the procedure outlined above, both $\operatorname{tr}(e)$ and $\det(e)$ can be expanded to second order:
\begin{equation}\label{tre}
\begin{aligned}
  \text{tr}(e)&=2+2\Phi+\partial_i\alpha^i+\Phi^2+\Phi\partial_i\alpha^i-\chi\varepsilon^{ij}\partial_i\alpha_j-\chi^2~~~~~~~~~~~~\\&+\mathcal{O}(\alpha^3,\alpha^2\Phi,\alpha\Phi^2,\Phi^3),
  \end{aligned}
  \end{equation}
\begin{equation}\label{dete}
\begin{aligned}  \det(e)&=1+2\Phi+\partial_i\alpha^i+2\Phi^2+2\Phi\partial_i\alpha^i+\frac{1}{2}(\partial_i\alpha^i)^2-\frac{1}{2}\partial_i\alpha^j\partial_j\alpha^i\\&+\mathcal{O}(\alpha^3,\alpha^2\Phi,\alpha\Phi^2,\Phi^3).
\end{aligned}
\end{equation}
The quantity to be evaluated is the expression inside the square root in Eq.~\eqref{action2}. A direct computation gives
\begin{equation}
    \sqrt{(\text{tr}(e))^2-4\text{det}(e)}=e^{\Phi}\sqrt{\left(-(\partial_i\alpha^i)^2+2\partial_i\alpha^j\partial_j\alpha^i-4\chi\varepsilon^{ij}\partial_i\alpha_j-4\chi^2  \right)},
\end{equation}
Let the expression under the square root be $A^2$. Then this equation reduces to a quadratic equation in $\chi$, from which the possible values of $A$ are constrained by $A^2 \le -(\partial_i\alpha^i)^2 + \partial_i\alpha^j\partial_j\alpha^i + \partial_i\alpha_j\partial^i\alpha^j$. After comparing various possible choices, the maximum value is selected. This choice has two important advantages: first, it allows an analytic evaluation; second, it ensures that the parameterized gravitational action reduces to the form in \cite{HiranoShigemori:2020,Hirano:2025tkq}. Accordingly,
\begin{equation}\label{A}
    A^2=-(\partial_i\alpha^i)^2+\partial_i\alpha^j\partial_j\alpha^i+\partial_i\alpha_j\partial^i\alpha^j=2\sigma_{ij}\sigma^{ij}.
\end{equation}
Substituting Eqs.~\eqref{phi},~\eqref{tre},~\eqref{dete} into Eq.~\eqref{action2},~one can verify that the final result is exactly Eq.~\eqref{parameterized action}.

\section{The solution of the Green's function of $\hat{O}_{ij}$}\label{appendixB}
In this appendix, the Green's equation of $\hat{O}_{ij}$ is solved, which reads as
\begin{equation}\label{Green equation}
\hat{O}_{ij}G^{jk}(x,x',y,t)=\delta^{~k}_{i}\delta^2(x-x').
\end{equation}
When \(x \ne y\), the Green's equation reduces to
\begin{equation}\label{reduced Green eq}
    \frac{1}{4\lambda}\delta_{ij}\Box G^{jk(0)}(x,x')=\delta_i^{~k}\delta^2(x-x'),
\end{equation}
whose non-trivial solution is
\begin{equation}
G^{jk(0)}(x,x')=\delta^{jk}\frac{2\lambda}{\pi}\ln\left(\frac{|x-x'|}{\varepsilon}\right).
\end{equation}
Based on this form, the solution is conjectured to take the form:
\begin{equation}\label{trial}
 G^{jk}(x,x',y,t)=G^{jk(0)}(x,x')+\mathcal{G}^{jk}(x,x',y,t).   
\end{equation}
Substituting the trial solution~\eqref{trial} into the Green’s equation~\eqref{Green equation} yields:
\begin{equation}\label{block}
\begin{aligned}
&\underbrace{\left(\frac{1}{4\lambda}-2t\delta^2(x-y)\right)\delta_{ij}\Box\mathcal{G}^{jk}(x,x',y,t)}_{\clubsuit}+\underbrace{\bigg(-8\lambda t\delta_{i}^{~k}\delta^2(x-y)\delta^2(x-x')\bigg)}_{\spadesuit  }\\
&+\underbrace{\bigg(-2t\hat{M}_{ij}G^{jk(0)}(x,x')\bigg)}_{\diamondsuit}+\underbrace{\bigg(-2t\hat{M}_{ij}\mathcal{G}^{jk}(x,x',y,t)\bigg)}_{\heartsuit}=0.
\end{aligned}
\end{equation}
A Fourier transform with respect to $x$ is performed on both sides. For clarity, the operations are carried out in blocks as indicated in Eq.~\eqref{block}. The Fourier transform is defined as 
\begin{equation}\tilde{\mathcal{G}}^{jk}(k,x',y,t)\equiv\mathcal{F}\big(\mathcal{G}^{jk}(x,x',y,t)\big)(k,x',y,t)=\int \text{d}^2x~\mathcal{G}^{jk}(x,x',y,t)e^{-ikx}.
\end{equation}
Here $\star$ denotes convolution. A direct calculation yields
\begin{equation}\label{clubsuit}
\begin{aligned}
\mathcal{F}(\clubsuit )&=-\frac{1}{4\lambda}\delta_{ij}k^2\tilde{\mathcal{G} }^{jk}(k,x',y,t)+2t\frac{1}{(2\pi)^2}\delta_{ij}e^{-ik\cdot y}\star k^2 \tilde{\mathcal{G} }^{jk}(k,x',y,t)\\
&=-\frac{1}{4\lambda}\delta_{ij}k^2\tilde{\mathcal{G} }^{jk}(k,x',y,t)+2t\frac{1}{(2\pi)^2}\delta_{ij}\int \text{d}^2\xi~\xi^2\tilde{\mathcal{G} }^{jk}(\xi,x',y,t)e^{-i(k-\xi)\cdot y}.
\end{aligned}
\end{equation}
Next,
\begin{equation}\label{spadesuit}
\mathcal{F}(\spadesuit)=-8\lambda t\delta^{~k}_i\delta^2(x'-y)e^{-ikx'}.
\end{equation}
Next,
\begin{equation}\label{diamondsuit}
\begin{aligned}
\mathcal{F}(\diamondsuit)&=-2t\left(\frac{\delta_i^{~k}}{(2\pi)^2}ik^\mu e^{-ik\cdot y}\star ik_\mu\left(\frac{-4\lambda e^{-ik\cdot x'}}{k^2} \right)-\frac{2\delta^{jk}}{(2\pi)^2} ik_{[i}e^{-ik\cdot y}\star ik_{j]}\frac{-4\lambda e^{-ik\cdot x'}}{k^2} \right)\\
&=-\frac{8t\lambda}{(2\pi)^2}\int \text{d}^2\xi\frac{e^{-i\xi\cdot(y-x')-ik\cdot x'}}{|k-\xi|^2}\left(\delta^k_{~~i}\xi\cdot(k-\xi)-2\delta^{jk}\xi_{[i}k_{j]}\right)\\
&=-\frac{8t\lambda e^{-ik\cdot y}}{(2\pi)^2}\int \text{d}^2p\frac{e^{ip\cdot(y-x')}}{p^2}\left(\delta^k_{~~i}(k-p)\cdot p+2\delta^{jk}p_{[i}k_{j]}\right)\\
&=\frac{4t\lambda}{\pi}\bigg( 2\pi e^{-ik\cdot x'}\delta_i^{~k}\delta^2(y-x')-\frac{ i e^{-ik\cdot y}}{|y-x'|^2}(-2\delta^{kj}k_{[i}(y-x')_{j]}+\delta_i^{~k}k\cdot(y-x')) \bigg).
\end{aligned}
\end{equation}
In the third step, the variable transformation $p=k-\xi$ has been performed. Next,
\begin{equation}\label{heartsuit}
\begin{aligned}
\mathcal{F}(\heartsuit)&=\frac{2t}{(2\pi)^2}\bigg(\delta_{ij}k^\mu e^{-ik\cdot y}\star k_\mu\tilde{\mathcal{G}}^{jk}(k,x',y,t)-2k_{[i} e^{-ik\cdot y}\star k_{j]}\tilde{\mathcal{G}}^{jk}(k,x',y,t) \bigg)\\
&=\frac{2t}{(2\pi)^2}\int \text{d}^2\xi ~\tilde{\mathcal{G}}^{jk}(\xi,x',y,t)e^{-i(k-\xi)\cdot y}\left(\delta_{ij}\xi\cdot(k-\xi)-2(k-\xi)_{[i}\xi_{j]} \right)
\end{aligned}
\end{equation}
Adding Eqs.~\eqref{clubsuit}--\eqref{heartsuit} together, several terms cancel, leaving the final equation as:
\begin{equation}\label{integral eq}
\begin{aligned}
-\frac{1}{4\lambda}k^2\tilde{\mathcal{G}}^{~k}_i(k,x',y,t)&+\frac{2te^{-ik\cdot y}}{(2\pi)^2}\int \text{d}^2\xi(\delta_{ij}k_\mu\xi^\mu+k_j\xi_i-k_i\xi_j)\tilde{\mathcal{G}}^{jk}(\xi,x',y,t)e^{i\xi\cdot y}\\
&-\frac{4\lambda it e^{-ik\cdot y}}{\pi|y-x'|^2}\bigg(-k_i(y^k-x'^k)+k^k(y_i-x'_i)+\delta_i^{~k}k\cdot(y-x')\bigg)=0.
\end{aligned}
\end{equation}
By applying the inverse Fourier transformation,~Eq.~\eqref{integral eq} can be recast as
\begin{equation}\label{self-referential eq}
\begin{aligned}
&\mathcal{G}_{ik}(x,x',y,t)-\frac{8\lambda^2 t}{\pi^2|y-x'|^2}\bigg( 2(y-x')_{[i}\partial_{k]}\ln\left(\frac{|x-y|}{\varepsilon}\right)+\delta_{ik}(y^\mu-x'^\mu)\partial_\mu\ln\left(\frac{|x-y|}{\varepsilon}\right)\bigg)\\
&-\frac{8\lambda t}{(2\pi)^3}\left.\bigg(2\partial_{[j}\ln\left(\frac{|x-y|}{\varepsilon}\right)\partial_{i]}^*\mathcal{G}^{j}_{~~k}(y^*,x',y)+\partial^\mu\ln\left(\frac{|x-y|}{\varepsilon}\right)\partial_\mu^*\mathcal{G}_{ik}(y^*,x',y)\bigg)\right|_{y^*\to y}=0.
\end{aligned}
\end{equation}
Here, the notation $\partial_i^* \equiv \frac{\partial}{\partial y_*^i}$ is used. This is a typical self-referential equation that can generally be expressed as a geometric series. One can verify that the formal solution to Eq.~\eqref{self-referential eq} is given by
\begin{equation}
\mathcal{G}_{ik}(x,x',y,t)=\frac{8\lambda^2 t}{\pi^2|y-x'|^2}\bigg( 2(y-x')_{[i}\partial_{k]}\ln\left(\frac{|x-y|}{\varepsilon}\right)+\delta_{ik}(y^\mu-x'^\mu)\partial_\mu\ln\left(\frac{|x-y|}{\varepsilon}\right)\bigg).
\end{equation}
All remaining divergences are removed via point-splitting regularization, which absorbs terms such as $\delta(0)$ and its derivatives into renormalization constants \cite{Hirano:2025tkq}. The final finite solution is thus given by the expression above.
\begin{eqnarray}
\mathcal{G}_1^{~1}(x,x',y,t)&=&\mathcal{G}_2^{~2}(x,x',y,t)=-\frac{8
\lambda^2t}{\pi^2}\frac{(x'-y)\cdot(x-y)}{|x'-y|^2|x-y|^2},\\
\mathcal{G}_1^{~2}(x,x',y,t)&=&-\mathcal{G}_2^{~1}(x,x',y,t)=-\frac{8
\lambda^2t}{\pi^2}\frac{\bigg((x'-y)\times(x-y)\bigg)_3}{|x'-y|^2|x-y|^2}.
\end{eqnarray}
Which corresponds precisely to Eqs.~\eqref{Green function},~\eqref{symmetric part} and \eqref{anti-symmetric part}.

\section{Proof of self-adjointness}\label{appendixC}
The Gaussian integral formula in the path integral is used, namely,
\begin{equation}\label{Guassian integral}
\int\mathcal{D}\alpha~e^{-\int \text{d}^2x~\alpha^i(x)\hat{O}_{ij}\alpha^j(x)+\int \text{d}^2x~ J(x)\cdot \alpha(x)}\propto \text{det}^{-\frac{1}{2}}(\hat{O}_{ij})e^{\frac{1}{4}\iint \text{d}^2x\text{d}^2x' J^i(x)G_{ij}(x,x',y,t)J^j(x')}.
\end{equation}
Here, $G_{ij}(x,x',y,t)$ is determined by Eq.~\eqref{Green equation}. For Eq.~\eqref{Guassian integral} to hold, the operator $\hat{O}_{ij}$ must satisfy two conditions: 
\begin{enumerate}
    \item $\hat{O}_{ij}$ must be invertible, which ensures that its determinant is non-zero, that is, a Green's function exists. This has already been proven in Appendix~\ref{appendixC}. 
    \item $\hat{O}_{ij}$ must possess symmetry, which guarantees that $\hat{O}_{ij}$ has a real spectrum and a complete orthonormal system of eigenfunctions. 
\end{enumerate}
The symmetry condition is verified below. Defining the inner product \begin{equation}\langle \psi,\phi \rangle \coloneqq \int \text{d}^2x ~ \psi(x) \phi(x)\end{equation} for arbitrary vectors $\psi^i(x)$ and $\phi^j(x)$, the condition to be shown is:
\begin{equation}\label{symmetry eq}
\left \langle \psi,\hat{O}\phi \right \rangle=\left \langle \hat{O}\psi,\phi \right \rangle  .
\end{equation}
Eq.~\eqref{symmetry eq} can be verified by integration by parts. Alternatively, it follows from the symmetry of the Green function~\eqref{Green function}. Indeed, setting\begin{equation}\phi^i(x)=\int \text{d}^2x' G^{ij}(x,x',y,t)(\hat{O}\phi)_j(x')\equiv \hat{O}^{-1}(\hat{O}\phi)^i(x).
\end{equation}
The first equality holds due to $\hat{O}$ is invertible, which means its null space is trivial.~Then
\begin{equation}\label{D4}
    \left \langle \hat{O}\psi,\phi \right \rangle =\left \langle \hat{O}\psi,\hat{O}^{-1}(\hat{O}\phi) \right \rangle =\left \langle \hat{O}^{-1}(\hat{O}\psi),\hat{O}\phi \right \rangle=\left \langle \psi,\hat{O}\phi \right \rangle.
\end{equation}
The second equality holds because Eqs.~\eqref{Green function},~\eqref{symmetric part} and \eqref{anti-symmetric part} give
\begin{equation}
    G_{ij}(x,x',y,t)=G_{ji}(x',x,y,t).
\end{equation}
By renaming the integration variables, one can verify the correctness of the second equality in Eq.~\eqref{D4} above. Therefore, Eq.~\eqref{Guassian integral} holds. Taking the external source $J^i(x)=ik^i\left(\delta^2(x-x_1)-\delta^2(x-x_2)\right)$ yields
\begin{equation}
   \int\mathcal{D}\alpha~ e^{-\int \text{d}^2x~\alpha^i(x)\hat{O}_{ij}\alpha^j(x)+ik\cdot \alpha_{12}}=Z_{\text{eff}}(t)e^{\frac{\lambda k^2}{\pi}S(x_1,x_2,y,t)}.
\end{equation}
This is how the second line of Eq. \eqref{calculation1} arises.

\section{Derivation of kernel \eqref{kernel}}\label{appendixD}
Starting with the integral representation of the Gamma function, the operator (denoted by $\hat{R}_{T\bar T}$) in Eq.~\eqref{operator of twopoint} can be written as
\begin{equation}\label{RTT}
 \hat{R}_{T\bar T}=\frac{1}{\Gamma(\Delta)^2} \int_0^\infty \int_0^\infty \text{d}x \text{d}y \, x^{\Delta-1} y^{\Delta-1} e^{-x-y}e^{\frac{2\lambda xy}{\pi}e^{\partial_\Delta}\partial_\Delta}.   
\end{equation}
Whereas,~the eigenvalue of $e^{\frac{2\lambda xy}{\pi}e^{\partial_\Delta}\partial_\Delta}$ in the Fourier basis $e^{ik\Delta}$ is:
\begin{equation}
    e^{\frac{2\lambda xy}{\pi}e^{\partial_\Delta}\partial_\Delta}e^{ik\Delta}=e^{\frac{2\lambda xy}{\pi}ike^{ik}}e^{ik\Delta}.
\end{equation}
Then its integral kernel in the Fourier basis is
\begin{equation}
    \tilde K_{T\bar T}(\Delta; \Delta') = \int_{-\infty}^\infty \frac{\text{d}k}{2\pi} e^{\frac{2\lambda xy}{\pi} ik e^{ik}} e^{ik(\Delta - \Delta')}.
\end{equation}
Hence,~the integral kernel $K_{T\bar T}(\Delta;\Delta')$ can be obtained through substituting $\tilde K_{T\bar T}(\Delta; \Delta')$ with the corresponding operator in Eq.~\eqref{RTT}.~Namely,
\begin{equation}
\begin{aligned}   
    K_{T\bar T}(\Delta;\Delta') &= \frac{1}{\Gamma(\Delta)^2} \int_0^\infty \int_0^\infty \text{d}x \text{d}y \, (xy)^{\Delta-1} e^{-x-y}\\
    &\times\int_{-\infty}^\infty \frac{\text{d}k}{2\pi} \exp\left(ik(\Delta-\Delta')+\frac{2\lambda xy}{\pi} ik e^{ik}\right).
\end{aligned}
\end{equation}
From the definite integral identity
\begin{equation}
\int_{0}^{\infty} \int_{0}^{\infty} x^{a-1} y^{b-1} e^{-px-qy+rxy}\text{d}x\text{d}y = \Gamma(a) \Gamma(b) p^{-a} q^{-b} {}_2 F_1 \left(a, b; 1; \frac{r}{pq}\right),
\end{equation}
the kernel $\hat{R}_{T\bar T}$ in Eq.~\eqref{kernel} follows as the Fourier transform of ${}_2F_1\left( \Delta, \Delta; 1; \frac{2\lambda}{\pi} i k e^{ik} \right)$.


\providecommand{\href}[2]{#2}\begingroup\raggedright\endgroup

\end{document}